\documentclass{egpubl}
\usepackage{eurovis2021}
\pdfoutput=1

\newcommand{\bstartnc}[1]{\vspace{1mm} \noindent{\textbf{#1}}}

\usepackage{xcolor}
\usepackage{xspace,xpunctuate}
\newcommand{\ie}{{i.e.,}\xspace}
\newcommand{\eg}{{e.g.,}\xspace}
\newcommand{\ea}{{et~al\xperiod}\xspace}

\newcommand{\etc}{{etc\xperiod}\xspace}

\SpecialIssuePaper         

\usepackage[T1]{fontenc}
\usepackage{dfadobe}  
\usepackage[normalem]{ulem}

\usepackage{cite}  
\BibtexOrBiblatex
\electronicVersion
\PrintedOrElectronic
\ifpdf \usepackage[pdftex]{graphicx} \pdfcompresslevel=9
\else \usepackage[dvips]{graphicx} \fi

\usepackage{egweblnk}

\title[Design Patterns and Trade-Offs in Responsive Visualization for Communication]%
      {Design Patterns and Trade-Offs in Responsive Visualization\\ for Communication}

\author[H. Kim, D. Moritz, \& J. Hullman]
{\parbox{\textwidth}{\centering 
Hyeok Kim$^{1}$\orcid{0000-0003-4340-4470}
Dominik Moritz$^{2}$\orcid{0000-0002-3110-1053}
Jessica Hullman$^{1}$\orcid{0000-0001-6826-3550}}
        \\
    {\parbox{\textwidth}{
        \centering $^1$Northwestern University
                   $^2$Carnegie Mellon University}
    }
}

\begin{document}

\maketitle
\begin{abstract}
Increased access to mobile devices motivates the need to design communicative visualizations that are responsive to varying screen sizes.
However, relatively little design guidance or tooling is currently available to authors.
We contribute a detailed characterization of responsive visualization strategies in communication-oriented visualizations, identifying 76 total strategies by analyzing 378 pairs of large screen (LS) and small screen (SS) visualizations
from online articles and reports.
Our analysis distinguishes between the \textit{Targets} of responsive visualization,
referring to what elements of a design are changed and
\textit{Actions} representing how targets are changed.
We identify key trade-offs related to authors’ need to maintain graphical density, referring to the amount of information per pixel,
while also maintaining the ``message’' or intended takeaways for users of a visualization.
We discuss implications of our findings for future visualization tool design to support responsive transformation of visualization designs, including requirements for automated recommenders for communication-oriented responsive visualizations.
\\


\end{abstract}  


\section{Introduction}\label{sec:intro}
Increased access to visualizations on mobile devices in contexts like online media~\cite{Fedeli2018,Lu2017} demands knowledge and tools for transitioning communicative visualization designs across display sizes. 
The process of designing for multiple display sizes, specifically focusing on transitioning larger screen (LS; \eg~desktop) views to mobile views, is often referred to as \textit{responsive visualization}~\cite{hinderman2015building,Andrews2018,hoffswell2020}.
In practice, a few simple responsive strategies are well known, such as proportional rescaling~\cite{bremer2019} or responsive layout specification~\cite{mobilevis}. 
However, many design challenges or trade-offs that arise in transitioning visualization designs for small screens (SS) remain difficult for authors to address.
For example, simple rescaling may cause overplotting of marks and make it difficult to select marks on small touch screens. 
Removing interactions reduces the content of a visualization in ways that might threaten its ability to convey the same message to mobile readers as the original did. 

Recent work~\cite{hoffswell2020} takes steps toward better support for authoring responsive visualization through a prototype visualization authoring system that enables authors to propagate design edits across different screen size versions of a visualization. 
However, the design space of responsive visualization strategies itself may be large, making it tedious to manually try out changes one by one. 
A deeper understanding of the design space of responsive visualization techniques--including a detailed characterization of what elements authors tend to add, remove, or change, how they do so, and what trade-offs motivate their choices--is a first step toward formalizing responsive visualization design knowledge to further support authors through automated design recommendations. 

Toward this goal, 
we first contribute a comprehensive summary of design strategies that authors currently use when creating SS versions of LS designs.
By comparing the LS and SS views of 378 public facing visualizations, we identify 76 design patterns for responsive visualizations.
Our analysis captures responsive visualization strategies (from LS to SS) in terms of \textit{Targets}, representing what is changed (data, encoding, interaction, narrative, references/layout) and \textit{Actions}, representing how the targets are changed 
(\eg~increase bin size, aggregate, reduce width, externalize annotations). 
Readers can explore our design strategies with illustration and descriptions on our online gallery\footnote{\url{https://mucollective.github.io/responsive-vis-gallery/} While we are not able to include screenshots due to copyright, readers can find our annotations for responsive design changes.}.

The second contribution of our analysis is to characterize key trade-offs in responsive visualization authoring.
We propose that the overarching design challenge in responsive visualization is a density-message trade-off where authors seek to balance goals of maintaining graphical density with those of preserving the message or intended takeaways of their work.
We observe that strategies addressing graphical density, layout, and interaction complexity often result in ``message loss'', where ``message'' captures a viewer's ability to recognize certain comparisons or relationships.
We identify different forms of message loss, including loss of information, interaction,  discoverability, concurrency of elements, and graphical perception.
We conclude by discussing the implications of our characterization of design patterns and key trade-offs for responsive visualization tool design.
\section{Related Work}\label{sec:rv}
\subsection{Needs for Responsive Visualization}
Responsive visualization involves several visualization scalability issues including display scalability and level of detail.
Display scalability is known as a key challenge in designing for visual analysis, referring to how well a visualization design scales for multiple device types with varying screen sizes and interaction methods~\cite{cook2005illuminating}.
Dealing with display scalability often requires more than simply rescaling to different screen sizes (\eg~everyday devices like desktops and smartphones for responsive visualization).
For example, scalability challenges arise from how well different viewing factors (\eg viewing distance~\cite{Isenberg2013}, chart size~\cite{Matkovic2002}) support presenting different levels of details.
To enhance scalability, prior work contributes algorithms for managing levels of detail through progressive refinement~\cite{Rosenbaum2009,Rosenbaum2012} or by limiting ``the number of visual entities''~\cite{Elmqvist2010}).
Visualization retargeting studies (\eg~Wu \ea~\cite{Wu2013}, Di Giacomo \ea~\cite{Giacomo2015}) provide algorithms for resizing charts while keeping visually salient information. 
Scalability concerns arise across various device types, such as scaling up desktop visualizations to wall-sized displays~\cite{Jakobsen2011,Jakobsen2013} and non-rectangular devices (\eg~circular tabletops~\cite{Vernier2002}, smart watches~\cite{Blascheck2019}).
In this study, we focus on two device types, LS (desktop/laptop) and SS (smartphones) devices as they are most commonly used devices for our scope of communicative visualizations.

The term responsive visualization draws an analogy to responsive web design~\cite{hinderman2015building}.
Hinderman~\cite{hinderman2015building} and Körner~\cite{korner2016learning} introduced responsive visualization techniques using D3~\cite{d3}.
Andrews~\cite{Andrews2018} demonstrated several responsive visualization techniques, including toggling fields on parallel coordinates and removing axes of a line graph.
Visualization designers (\eg~Bremer~\cite{bremer2019} and Ros~\cite{mobilevis}) have also described design strategies for visualization on both mobile and desktop, including repositioning, rescaling, stacking, zooming, and immobilizing.

The need for responsive visualization stems from the physical and contextual differences between various device types~\cite{Chittaro2006}.
First, the smaller screen size and portrait aspect ratio of SS devices require different visualization specifications, primarily because visual marks and letters need a certain minimum pixel-space difference (\eg~size, position, hue, \etc) to be recognized.
Second, while LS devices receive inputs through keyboard and pointing devices, SS devices usually use touch interfaces. 
Because touch interactions are less accurate on mobile devices (\eg~due to the fat-finger problem~\cite{Lee2012} and a limited touchable screen area~\cite{Lehmann2018}), interactions often must be altered. 
Third, the reduced computational power of SS devices creates problems rendering dynamic and complex visual representations and interactions.

Designers should take contextual characteristics, such as the conditions, purpose, and length of use into account in responsively transforming a visualization. 
People often use SS devices under conditions that make it hard to focus (\eg~walking~\cite{Conradi2017} or using them with other devices~\cite{Google2016}).
People are likely to use SS devices for simpler purposes (\eg~instant messaging or pickups) for shorter amounts of time~\cite{MacKay2019}.
These contextual differences between LS and SS imply that authors often need to tune their SS visualization according to a more focused subset of their intentions (\eg~simplifying or emphasizing elements in terms of importance) to prevent them from overwhelming readers.

\subsection{Responsive Visualization Techniques}
Prior work has inspected various design strategies for visualizations on SS devices, such as using different layout styles (linear vs. radial)~\cite{brehmer2019}, comparing animation and small multiples~\cite{brehmer2019comparative}, connecting related points in scatterplot~\cite{radloff2011supporting}, or rectangularizing radiar views~\cite{Chhetri2015}.
Open-source APIs like Google Charts~\cite{googlechart} use rules to generate SS views, including using ellipsis (``...'') for overflowing labels, and removing overlapping labels.
Our work outlines a larger design space of strategies for responsive visualization and considers trade-offs between information density and the preservation of intended ``messages'' or takeaways in responsive visualization to inform more sophisticated forms of software support for authors.

Beyond the aforementioned empirical studies, recent visualization research has contributed software to support responsive visualization.
Leclaire \ea~\cite{leclaire2015} offer R3S.js, a JavaScript library that manages JS events, tooltips, media queries, and axes.
Hoffswell \ea~\cite{hoffswell2020} present a prototype authoring system for responsive visualization that supports view concurrency and edit propagation within multiple views.
Similar to our work, to inform their tool, they describe responsive visualization techniques in a corpus of 231 LS-SS visualization pairs using five predicates (resize, reposition, add, modify, and remove). 
Our work extends their taxonomy considerably by detailing 76 strategies describing \textit{how} authors add, modify, and remove elements. 
Wu \ea~\cite{wu2020mobilevisfixer} provide MobileVisFixer, a reinforcement learning-based approach to translating a non-responsively designed Web visualization to a mobile-friendly view.
They focus on strategies for adjusting font size, axes, ticks, and margins and adopt related cost functions based on heuristics like ``out of the viewport,'' ``unreadable font-size.''
Their approach is limited to addressing a narrow set of issues that arise from simple transformations of an LS to an SS view (\eg~reducing ticks, breaking text line) but which can be automated.
Our goal is instead to understand the larger space of design strategies that can be used in responsive visualization designs.

\section{Responsive Visualization Design Patterns}\label{sec:study1}

\begin{figure*}[t]
    \centering
    \includegraphics[width=\textwidth]{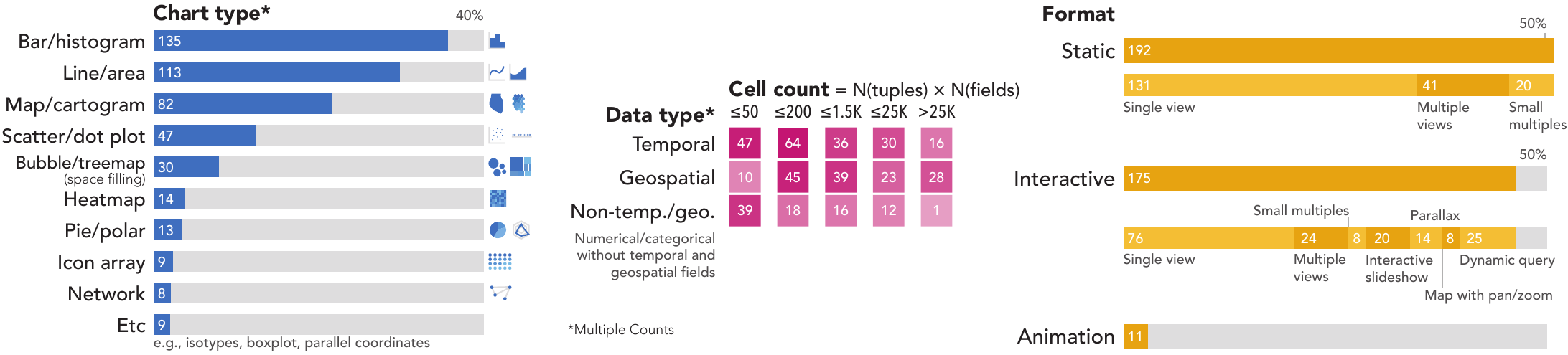}
    \caption{Properties of our visualization sample. We reconstructed cell count by multiplying the number of tuples (records) and the number of fields (columns) and grouped cell sizes by k-means clustering (k = 5).}
    \label{fig:sample}
\end{figure*}

Based on our qualitative analysis of large screen (LS) and small screen (SS) versions of 378 visualizations intended for communication, we characterize design strategies for responsive visualization and categorize them in terms of \emph{Targets} and \emph{Actions}.

\subsection{Methods}\label{sec:study1methods}

\subsubsection{Sample collection} 
We collected a sample of 378 \textit{pairs of} LS and corresponding SS visualizations (756 total visualizations) from the media (\eg~news outlets), data-driven reports from global organizations, and blog posts about responsive visualization. 
We first collected pairs of visualizations from 104 data-driven news articles containing visualizations from the \textit{New York Times} (NYT) and the \textit{Wall Street Journal} (WSJ)'s yearly galleries of data visualization articles (2016-2017)~\cite{nyt2016,nyt2017,wsj2016,wsj2017}.
We included all articles with abstract data visualizations that map numerical and/or categorical data to visual variables and excluded illustration and photography-based articles. From this set, we obtained 280 pairs of visualizations.
To this set we added visualizations from 57 additional articles and visualization projects from international organizations (OECD, UNESCO), visualization authors' blog posts (\eg~Bremer~\cite{bremer2019}), \textit{MobileVis} gallery~\cite{mobilevis}, and \textit{Scientific American}, which provided both LS and SS views (98 more pairs of visualizations).
\autoref{fig:sample} illustrates the properties of our sample. 
We provide the full list in our interactive gallery.
The size and diversity of sources in our sample suggest that it should offer a reasonable, albeit not comprehensive, snapshot of the design space for communication-oriented responsive visualization design.

\subsubsection{Analysis} 

To characterize design strategies,
two authors and an external coder iteratively coded differences between the LS and SS visualizations in each pair using methods from grounded theory~\cite{Corbin1990}.
We started with open-coding~\cite{lofland1971} to build up a large set of descriptions of differences between LS and SS versions of visualizations (\eg~add highlighting, remove an interaction feature).
We then made several additional coding passes, grouping observations made from different pairs into single, recurrent strategies and returning often to examine the sample visualizations to confirm that a strategy was in fact the same. 
This process resulted in 76 design patterns or strategies.
We observed that each of these strategies could be further distinguished by the \textit{Target} of the change, representing what type of visual or design element was changed (\eg annotations, data, encodings) and the \textit{Action} describing the form of the change (\eg removing, highlighting, increasing).
Finally, we developed higher level groupings of Targets and Actions shared across strategies, respectively.
This analysis distinguished five categories of Targets (Data, Encoding, Interaction, Narrative, and References/Labels) and five categories of Actions (\eg~Recompose, Rescale, Transpose, Reposition, and Compensate). 
We tabulated counts of different strategies observed across our sample and their co-occurrence in \autoref{sec:patterndistribution}.

\begin{figure*}[!ht]
\minipage{0.32\textwidth}
  \includegraphics[width=\columnwidth]{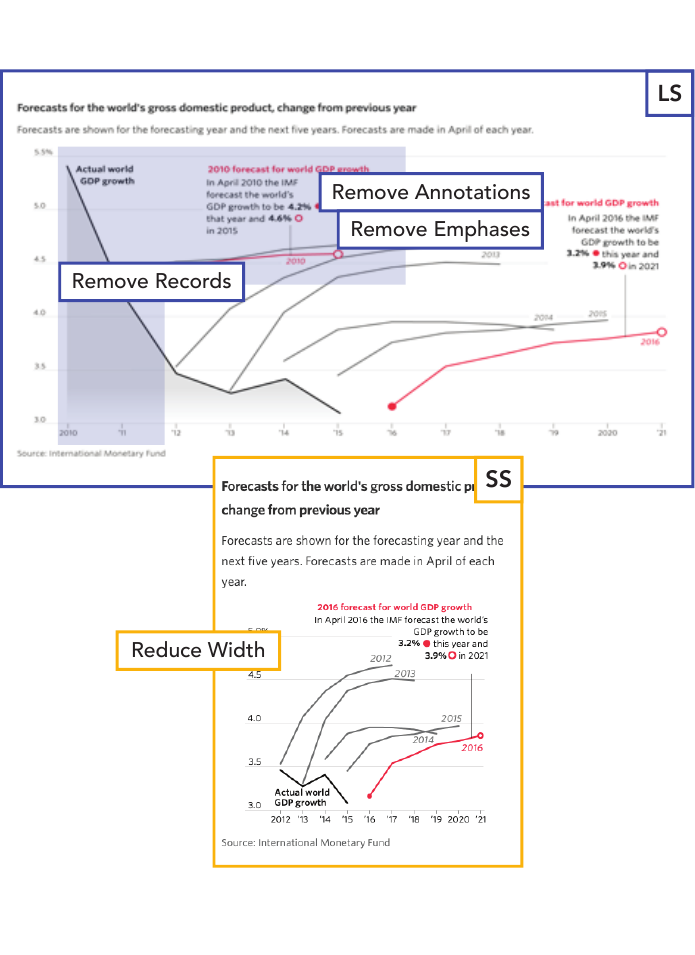}
    \caption{Screenshots of \textit{Bond Yield}'s LS and SS view pair that illustrates \textit{remove records}, \textit{remove annotations}, \textit{remove emphases}, and \textit{reduce width}. Blue highlights indicate parts of the LS view that are removed in SS.}
  \label{fig:ex_bondyield}
\endminipage\hfill
\minipage{0.32\textwidth}
  \includegraphics[width=\columnwidth]{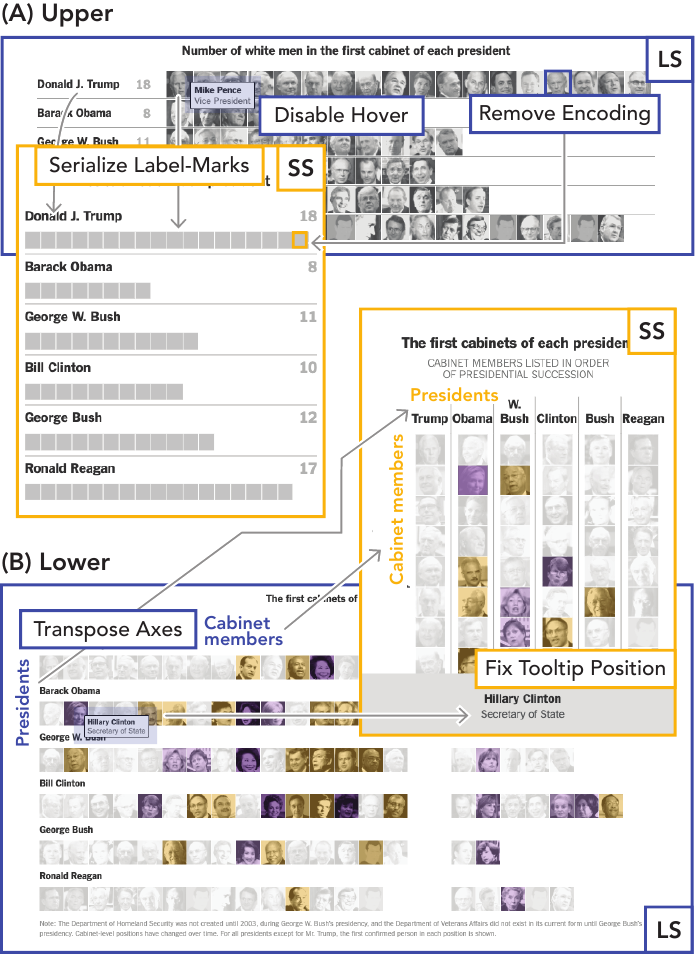}
    \caption{Screenshots of \textit{U.S. Cabinet}'s LS and SS view pair that demonstrates \textit{disable hover interaction}, \textit{remove encoding}, \textit{serialize label-marks}, \textit{transpose axes}, and \textit{fix tooltip position}.}
  \label{fig:ex_uscabinet}
\endminipage\hfill
\minipage{0.32\textwidth}
  \includegraphics[width=\columnwidth]{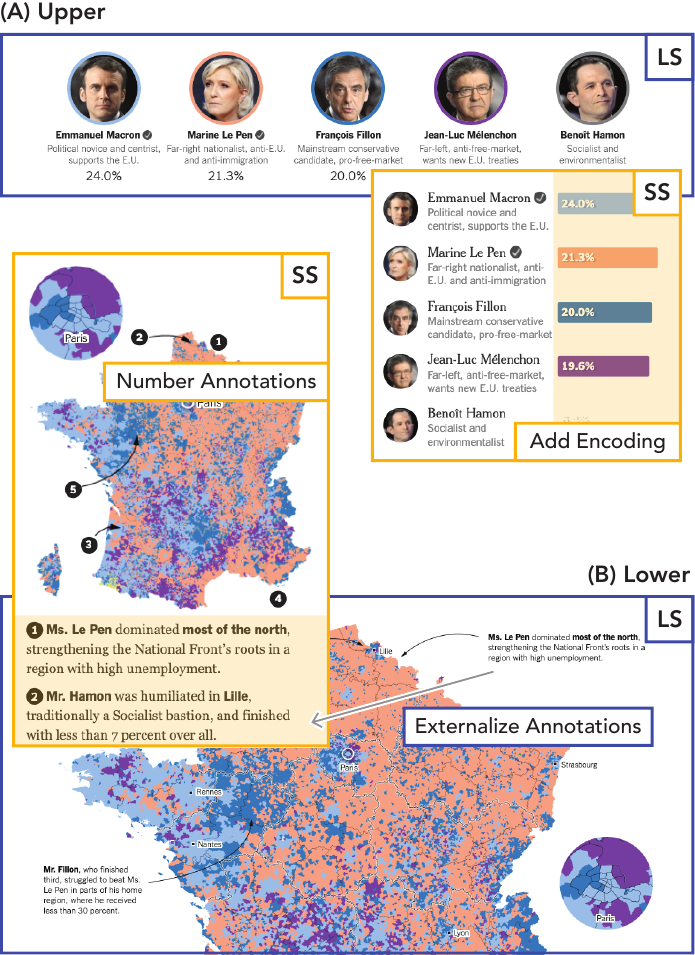}
    \caption{Screenshots of \textit{French Election}'s LS and SS view pair that demonstrates \textit{add encoding}, \textit{externalize annotations}, and \textit{number annotations}. Yellow highlights indicate parts that are added or repositioned in SS.}
  \label{fig:ex_frenchelection}
\endminipage
\end{figure*}

\subsubsection{Preliminary Survey of Authors}\label{sec:survey}
To supplement our analysis of examples, we surveyed 19 visualization authors with experience in designing responsive visualization (average 4.8 years), who we solicited through a posting on social media. 
We asked about their typical process of designing a responsive visualization (\ie~starting from an LS view, from an SS view, or designing both simultaneously) and how many times they consider SS views in their design process (less than 10\% of the time, less than half the time, about half the time, more than half the time, more than 90\% of the time). 
Next, we asked them to rank seven design guidelines for responsive visualization. The guidelines were informed by prior work on responsive visualization and our initial analysis, like `maintaining the main takeaways' and `maintaining the information density.'
We included open-ended questions to elicit any ``rules of thumb'' they used and difficulties they faced when authoring visualizations for mobile screens. 

Eleven authors (58\%) described creating SS visualizations after designing LS views, similar to the findings of a recent interview study~\cite{hoffswell2020} with five authors. 
As a result, we default to describing design strategies as transformations of an LS view in presenting our analysis.
Yet, the different Action strategies (\autoref{sec:dim2}) we identify are invertible, so this direction is primarily a communication mechanism rather than a property of our analysis.
When asked to rank different possible guidelines for responsive visualization, authors ranked ``maintaining takeaways,''  ``maintaining information,'' and ``changing the design to acknowledge greater interaction difficulty on an SS'' as most important. 
More than half (10, 53\%) of the authors described strategies they used and/or concerns they had in adjusting information density for smaller screens (\eg \textit{``Step by step information reveal rather than showing everything at once''} (P15), \textit{``Creating a similar experience without overwhelming the user''} (P18)). 
Such statements informed our identification of important trade-offs in responsive design.
Full survey questions and responses are provided in supplementary material (\url{https://osf.io/zrqfy/}).

\subsection{Examples}\label{sec:examples}
We provide three examples of responsive visualization to introduce the reader to key design patterns.

\subsubsection{\textit{Bond Yield} - Data, Annotation, and Size}\label{sec:bondyield}

\textit{Bond Yield}\footnote{\url{https://www.wsj.com/graphics/how-bond-yields-got-this-low/}} illustrates strategies of information removal from an LS to an SS view, and consequent changes in emphasis. 
The area mark on the left of the LS view in \autoref{fig:ex_bondyield} expresses observed world GDP growth from 2010 to 2015.
The SS view omits the grey area mark as well as two line marks representing five-year forecasts of GDP growth rate.
In the LS view, the omitted part had served to show that GDP growth rate projections were higher before the actual growth rate plummeted.
The authors retained lines in the SS view that show a more recent decrease in the International Monetary Fund's GDP growth rate forecasts.
Data records for the years 2010 and 2011 have been removed (\textbf{remove records}), resulting in further changes to axes, annotations, and emphases.
The scales of the \textit{x}-axis (years) and \textit{y}-axis (forecasted GDP growth) are consequently altered. 
The annotation and emphasis (in red and boldface) for the forecast of 2010 observed GDP growth have been omitted from the SS view (\textbf{remove emphases} and \textbf{remove annotations}). 
Additionally, the relative width of the SS view is slightly reduced (\textbf{reduce width}), compared to its height relative to the LS view.

Two interrelated intentions may be behind these changes. 
First, the authors may have wanted to avoid an overly dense display caused by placing two long annotations close to each other. 
Second, they may have intended to support a more glanceable reading of the visualization by reducing the number of key points.

\subsubsection{\textit{U.S. Cabinet} - Encoding and Tooltips}\label{sec:uscabinet}

\textit{U.S. Cabinet}\footnote{\url{https://nyti.ms/2jSp3WT}} compares race and gender ratios in recent U.S. Cabinets, and demonstrates changes to visual encodings and interactions.
In \autoref{fig:ex_uscabinet}, the LS views of both the \textit{upper} and \textit{lower} visualizations share several similarities.
They use the same encoding: a mapping of images of cabinet members' faces as bars. 
When the viewer hovers over each image in these LS views, a tooltip appears and shows that member's name and role. 
However, the \textit{upper} and \textit{lower} visualizations exhibit different responsive transformations in terms of encodings and tooltips.
First, in the \textit{upper} visualization (\autoref{fig:ex_uscabinet}A), the authors omitted the images of faces and tooltips from the SS view (\textbf{remove encoding}-a nominal variable and \textbf{disable hover interactions}, respectively).
Moreover, the labels and marks are serialized (\textbf{serialize label-marks}).
In the \textit{lower} visualization (\autoref{fig:ex_uscabinet}B), the axes are transposed from \textit{y} (presidents) $\times$ \textit{x} (Cabinet members) to \textit{y} (Cabinet members) $\times$ \textit{x} (presidents) (\textbf{transpose axes}). 
Also, the images of faces and the tooltip are preserved. 
However, in the SS view the position of tooltips is fixed to the bottom of the screen (\textbf{fix tooltip position}), while on LS, the tooltip is shown close to the corresponding image of a face (\ie where it is triggered).

A rationale behind these decisions may be the role of each visualization in the article's narrative.
The \textit{upper} visualization shows one aspect of the data (white males), while the \textit{lower} one provides a more comprehensive view including more variables (gender and race).
Instead of maintaining the same design in both the \textit{upper} and \textit{lower} views, which might result in high visual density, the authors of \textit{U.S. Cabinet} may have decided to simplify the \textit{upper} visualization while transposing the axes to fit the \textit{lower} visualization to the portrait aspect ratio. 

\subsubsection{\textit{French Election} - Addition and Compensation}\label{sec:frenchelection}

\textit{French Election}\footnote{\url{https://nyti.ms/2pbI1uD}}, a map-based static visualization of results of the French 2017 presidential election, illustrates strategies of adding an encoding and compensating problems caused by another strategy.
The \textit{upper} visualization in the LS view of \autoref{fig:ex_frenchelection}A uses a color encoding for the borders around the images of the candidate's faces (a nominal variable), but does not visually encode numerical data (the total vote shares of candidates), instead providing them as text.
However, the SS version encodes the total vote shares of candidates using \textit{x} position, resulting in a new bar graph (\textbf{add encoding}-a continuous variable).
A possible intention behind this decision might be to ensure that the viewer perceives the values on SS.

The choropleth map in the \textit{lower} visualization (\autoref{fig:ex_frenchelection}B) shows the distribution of the winners across France. 
Because of the discrepancy in population density between urban and rural areas, the predominant color on the map suggests a ranking that conflicts with the election outcome (\ie the pink candidate loses the election).
The authors rely on annotations to prevent misunderstandings in the LS view. 
However, showing these annotations on SS at similar positions is unlikely to fit on the screen, so the authors moved the annotations out of the choropleth (\textbf{externalize annotations}).
Presumably, to help readers locate the annotations to the map without background knowledge in French geography, the authors placed numbers in the original positions as a compensation method (\textbf{number annotations}).

\subsection{Design Patterns}\label{sec:patterns}
\begin{figure}[t]
    \centering
    \includegraphics[width=\columnwidth]{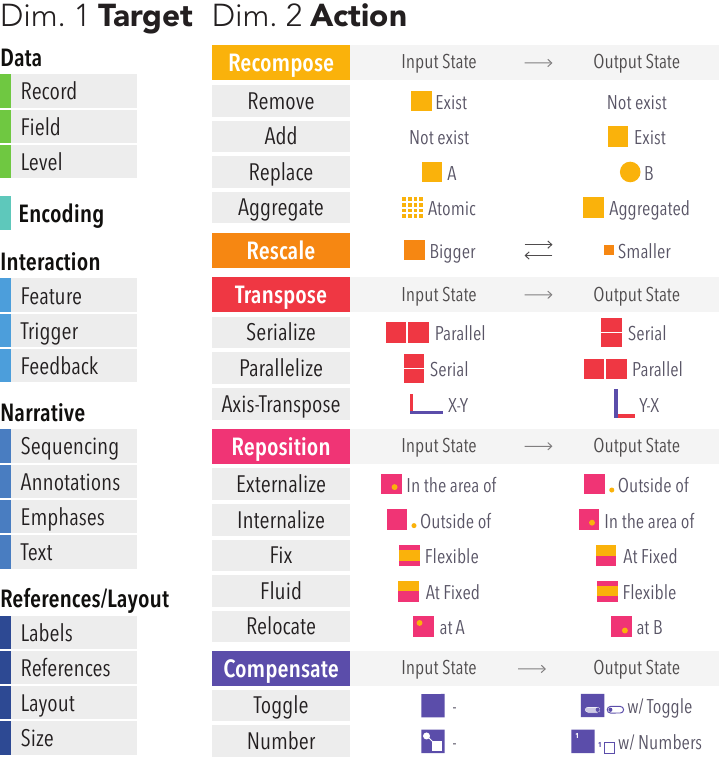}
    \caption{The dimensions of design patterns for responsive visualization.}
    \label{fig:ds_dimensions}
\end{figure}

\begin{figure*}[t]
    \centering
    \includegraphics[width=\textwidth]{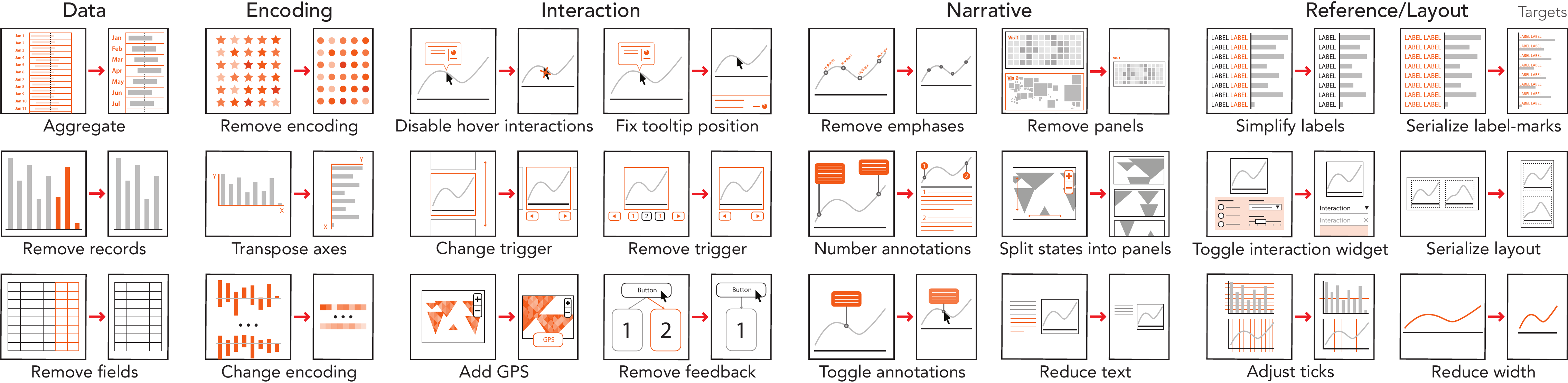}
    \caption{Examples of design patterns for responsive visualization, grouped by the target of the change (columns). The left and right side for each pattern denote LS and SS views, respectively. Orange is used to highlight those elements that change from LS to SS.} 
    \label{fig:pictogram}
\end{figure*}

Our characterization of design patterns for responsive visualization distinguishes two dimensions of design decisions: (1) the \textit{Target} (capturing what is changed from LS to SS), and (2) the \textit{Action} (capturing how the target is changed from LS to SS).
An overview of these two dimensions is shown in \autoref{fig:ds_dimensions}, and a sample of design patterns is illustrated in \autoref{fig:pictogram}.
On our explorable online gallery, we provide a design guide in the form of pictograms and descriptions of the entire set of patterns.
In the rest of this paper, we refer to visualization example articles in our interactive gallery for referenced strategies as E\#\footnote{This is reference to each example `article' that often has multiple sample visualizations, and the numbers are not consecutive.}.

\subsubsection{Targets--What Elements are Changed}\label{sec:dim1}
The Target dimension consists of five types of entities that can be transformed in creating SS designs: data, encodings, interaction, narrative, and references and layout.
The \textit{data} category includes records (or rows or tuples), fields (or columns), and levels of hierarchy (or nesting).
Transformations applied to data visualized in an LS view typically result in visible changes in an SS visualization: changing the number of \textit{records}, for example, can change the number of marks (\eg~line marks omitted in \textit{Bond Yield}), as can changes to \textit{levels of hierarchy} (\eg changing from showing daily measurements to monthly), while changes to what data \textit{fields} are shown typically result in encoding changes (\eg~detail/image encoding removed in \textit{U.S. Cabinet}).
Changes to an \textit{encoding} include switching a visual channel for showing a field (E139-size to length). 
The removal of an encoding often results from either removing a data field from the LS source data (E1-a nominal variable on texture, E209-a continuous variable on hue, E236-continuous variables on position) or eliminating a redundant encoding from the LS view (\textit{Bond Yield}-area under line, E15-hue).

\vspace{3mm}
The \textit{interaction} category describes targets related to a supported interaction in the LS view, including a feature, trigger, or feedback mechanism.
An interaction \textit{trigger}(s) refers to how a viewer provides input to interact and \textit{feedback} refers to the outcome of the interaction conveyed to the viewer.
The \textit{feature} subcategory refers to composites of interactions that realize a given functionality.  
For example, if a search feature receives user input via a text input box and an option list on LS, and the text input box is removed on SS, then this is a change in trigger (E150).
In our sample, we observed authors detached button triggers for zooming (when dragging interaction is available) (E14) and replaced a list of buttons with an option box (E139).
However, if the search interaction functionality is disabled on SS, we refer to it as removing a feature (E153).
Authors in our sample omitted various features including sorting (E114), filtering (E150), and map browsing (E226).
As illustrated in the \textit{U.S. Cabinet} example, authors can further remove interaction features (tooltip) after removing a data field (detail).

The \textit{narrative} category concerns sequence of information and authors' explicit messaging (annotation, emphases, text), inspired by prior work in narrative visualization~\cite{Segel2010,Hullman2011,hullman2013deeper,hullman2013contextifier}. 
\textit{Sequencing} deals with the existence of and methods for transitioning between multiple panels and states in a visualization.
Panels refer to multiple views existing concurrently (\eg in a poster style layout~\cite{Segel2010}), and states refer to multiple views sequenced or manipulated within the same panel (\eg interactive slideshow). 
Changes made to sequencing can involve interactions when the sequencing method relies on related interactions (\eg~themed/numbered tabs-\textit{split states into panels}).
Sequencing has to do with how viewers ``move'' from one element to another (\eg~a fixed order by position or interactive tabs, or a random order by panning or zooming)
whereas layout is more about how elements are placed on a page (\eg~horizontally versus vertically).
\textit{Annotations} and \textit{emphases} are often associated with important data points or ranges.
We use \textit{text} to refer to sentences or paragraphs that appear along with a visualization and summarize it (E21-adding a summary sentence).
Finally, the \textit{reference and layout} category includes labels and references to help the viewer read the encodings and how visualizations and surrounding elements like text are laid out in a display. 

\subsubsection{Action--How Targets are Changed}\label{sec:dim2}
Actions transform responsive visualization Targets in an LS view for an SS view. Hoffswell~\ea~\cite{hoffswell2020} described five high-level action categories (resize, reposition, add, modify, remove). 
We extend their understanding of an \textit{Action} dimension by defining action subcategories as functions with input and output states. 
We adopt this framing because it makes it possible to conceive of the inverses of the functions in a mobile-first design context. 
For instance, the inverse of \textit{externalize} is \textit{internalize}, and the inverse of \textit{relocate} is itself (\ie~un-relocating a target is another relocation of it).

The Action dimension consists of five categories: recompose, rescale, transpose, reposition, and compensate. 
The \textit{recompose} category involves actions that change the existence of a target, including remove, add, replace, and aggregate. 
\textit{Remove} actions refer to removing a target in the LS for the SS design. 
For example, a \textit{remove fields} pattern describes the removal of data fields, which leads to a concurrent removal of an encoding (\eg~\textit{U.S. Cabinet}-images of faces).
Authors in our sample often omitted hovering interactions for short labels (E19, E126) or tooltips (E226, E128) with corresponding hover highlight removed or maintained.
Complex interaction features are often removed, formulated as the pattern \textit{Disable X}, where \textit{X} can be one of various interactions, such as hypothesis (E13), search (E153, E222, E227), and filter (E2, E138, E150). 
An \textit{add} action inserts a target on SS that does not exist on the LS view. 
Small targets, such as a call-out line (E267), or legend (E222), are occasionally added for SS views. 
Sometimes, more elaborate targets, such as summary text for fast reading (E21), a location finder for simplified interaction (E2), and context views for reduced focus views (E126, E202) are added. 
Remove and add actions are invertible. 
We observed a few instances of \textit{replace} actions, referring to strategies that substitute a target in LS with another target in SS.
For example, \textit{change measurements} refers to a transformation of data values to encode (\eg~from mapping raw values to mapping ranks; E18).
An \textit{aggregate} pattern reduces lower level values in a given data set to higher level aggregates using various aggregate functions (\eg~sum-E45, mean, count).

\textit{Rescale} actions change a target from a bigger state to a smaller state or vice versa.
For example, a \textit{reduce width} pattern reduces the width of a visualization relative to the height, resulting in a narrower aspect ratio. 
A \textit{simplify labels} pattern shortens labels through a predefined mapping (E46-1980 to `80, E8-January to J) while an \textit{elaborate labels} pattern refers to detailing labels (E19) when the context for short labels is not concurrently visible.

\textit{Transpose} actions change the orientation of targets. 
\textit{Serialize} means placing two or more parallelly arranged elements on LS in a vertically serial order on SS.
Two or more panels, a pair of a visualization and a passage of text, or an interaction widget and a visualization are often serialized (\textit{serialize layout}). It was one of the most frequent strategies in our sample.
Within a visualization, labels and marks were frequently serialized (\textit{serialize label-marks}, E43, E59). 
As the inverse of serialize, \textit{parallelize} refers to placing two or more serially arranged elements on LS in a horizontally parallel order on SS. 
This was often applied to legends (E1) and labels (E41) in our sample. 
An \textit{axis-transpose} action exchanges \textit{x}- and \textit{y}-axes in charts with position encoding channels (\textit{transpose axes}, \textit{U.S. Cabinet}, E138) or a systematic layout of an interaction widget (\textit{transpose interaction widget}, E141).

The \textit{reposition} category refers to altering the position of targets, including externalize-internalize, fix-fluid, and relocate. 
When labels, legends, and annotations are placed close to corresponding visual marks, they are often \textit{externalize}d from visualization to reduce visual density (\textit{externalize labels}, E267; \textit{legends}, E116; \textit{annotations}, E262, E268). 
To the contrary, when small targets are placed outside of a visualization, they may be \textit{internalize}d (incorporated in the visualization) for effective use of space (\textit{incorporate labels}, E6, E207; \textit{internalize legends}, E116, E158). 
\textit{Fix} and \textit{fluid} actions refer to constraints on the arrangement of targets. 
For example, a tooltip for details-on-demand usually appears close to the corresponding data point when hovered on LS.
When a tooltip is big and likely to hide the chart on SS, the tooltip is often fixed at a particular position on screen (frequently at the bottom; \textit{fix tooltip position}, E1, E129, E204). 
Similarly, a text message that interactively appears on LS may be consistently displayed on an SS view (\textit{unhide text}, E222). 
Authors can fix sequencing by giving a strict viewing order, for example \textit{splitting explorable states into static panels} (E125, E132).
In contrast, a \textit{fluid} action refers to when elements are arranged in a fixed grid for on LS, but that grid no longer exists in SS.
For example, small multiples (E6, E16) and icon arrays (E19) are often rearranged following the screen width on SS (\textit{fluid small multiples} and \textit{fluid layout}, respectively). 
Authors at times \textit{relocate} targets by \textit{relocating annotations} (E20, E119) or \textit{moving marks} like non-contiguous territories in a map (E159) to vacant areas in a chart. 

Finally, the \textit{compensate} category involves techniques used to compensate for the loss of information. 
When it is difficult to arrange labels (E133) or legends (E209) due to limited screen space, authors can prevent losing the information by \textit{toggling} them. 
For example, it is possible to toggle an axis (\ie a data field) in a parallel coordinate plot (E201) with multiple axes. 
Another compensation technique is \textit{numbering}, which places numbers at the original positions of externalized targets on LS (\textit{French Election}). 

\begin{figure}[t]
    \centering
    \includegraphics[width=1\columnwidth]{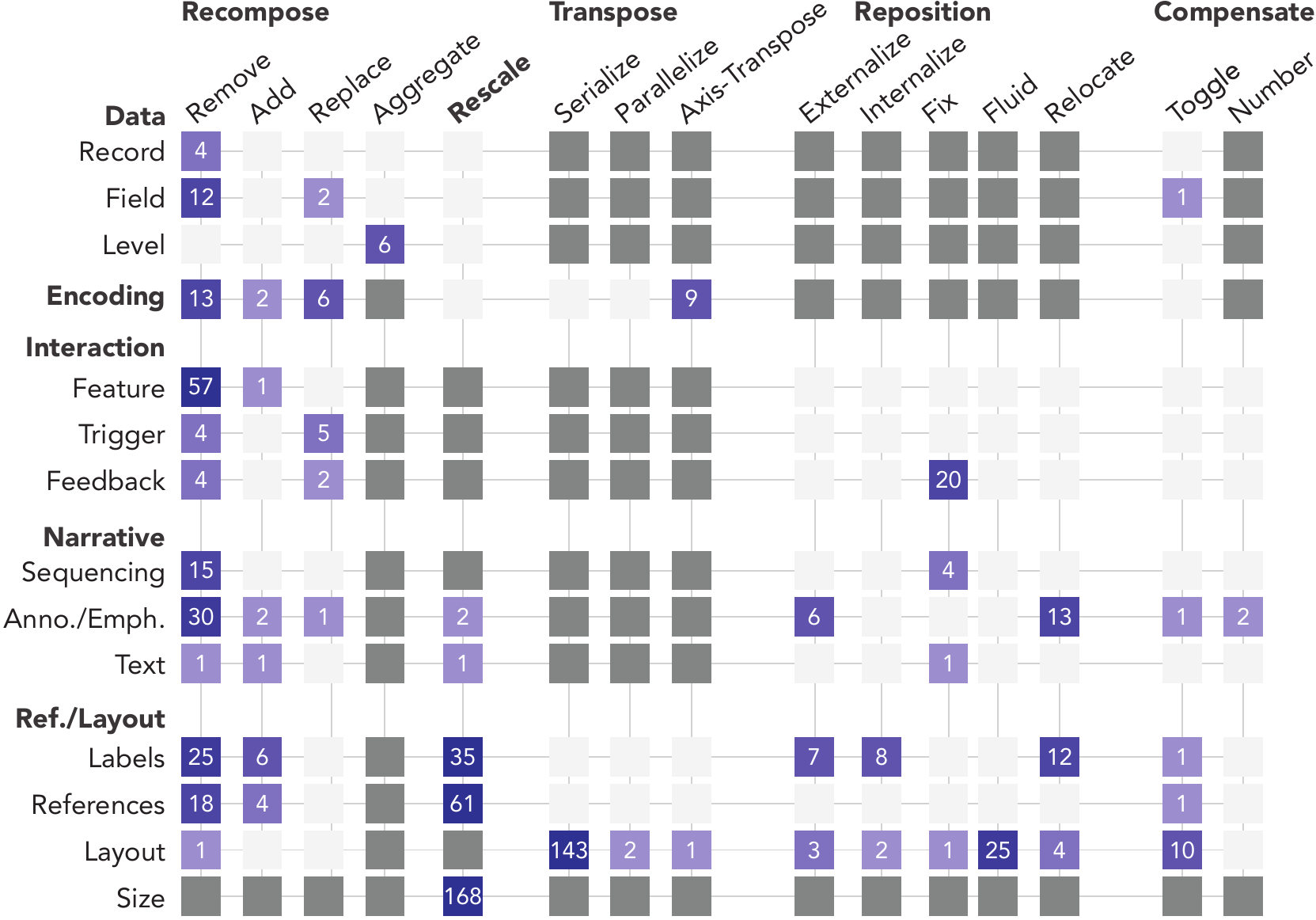}
    \caption{The distribution of responsive visualization strategies observed in our sample. Each strategy includes a Target (rows) and Action (columns), each of which we further categorized (bold labels). The gray-shaded cells denote combinations of target and action that we have not observed, and the dark gray cells indicate impossible combinations by definition. `Ref.' stands for `references.' }
    \label{fig:ds_dist}
\end{figure}

\subsubsection{Distribution of Responsive Visualization Strategies}\label{sec:patterndistribution}

As summarized in \autoref{fig:ds_dist}, authors applied responsive transformations most frequently to Reference/Layout targets, followed by Interaction and Narrative targets. 
Transforming Data and Encoding targets was less frequent, though multiple authors employed strategies like removing data fields and removing, transposing, or replacing encodings.
Overall, \textit{rescale} and \textit{remove} actions were most commonly used, with reducing size being most common, followed by \textit{transpose} actions specifically involving serializing layout.
We suspect that the automatability of these frequent strategies plays a role in their popularity. 
Authors may try to avoid substantive changes between LS and SS views, which require greater effort to make.
That authors are exploring various types of more complex transformations, even if not in great frequency, suggests that alternatives to canonical simple responsive transformations can be preferable.

The matrix format of \autoref{fig:ds_dist} is used for layout purposes and does not indicate that every combination of target and action is possible. 
Combinations of targets and actions that we did not observe may suggest new responsive visualization design techniques to explore. 
For instance, authors could add a sequencing method (\eg~from small multiples to an interactive slideshow) or could fix labels if an LS view has an extensive table format (\eg~freezing head columns).
However, some combinations of target and action are not possible by definition.
For example, authors cannot externalize data records or parallelize data fields because \textit{reposition} and \textit{transpose} actions are spatially defined while \textit{data} targets are not.
To the contrary, authors cannot aggregate interaction or layout because aggregation is a data-specific action although it can initiate downstream changes in other targets like labels.
\vspace{1mm}
\section{Trade-offs in Responsive Visualization}\label{sec:tradeoff}
Several insights from our analysis (\autoref{sec:study1}) suggest that responsive visualization design is characterized by a set of trade-offs among competing goals. 
First, different authors use different strategies to resolve seemingly similar problems, implying that a single ``best'' solution may not exist for many situations~\cite{rittel1973,buchanan1992}. 
Second, the \textit{compensate} actions that we observed imply trade-offs by suggesting that design strategies aimed at addressing one problem may result in other problems that need to be addressed.

Our survey of responsive visualization authors (\autoref{sec:survey}) indicates that authors may see maintaining the message of their visualization while adjusting information density (\ie~the amount of information per pixel) for SS devices as a central challenge in their work.
We identify a set of trade-offs in responsive design related to an overarching trade-off between information density and preserving a visualization's intended messages, which we conceptualize as a viewer’s ability to recognize certain comparisons or relationships in data. 
We describe three forms of information density problems that arise in transitioning designs from LS to SS views, and five distinct types of losses describing ways in which intended messages or takeaways can be lost in attempts to address these problems.

\begin{figure}[b]
    \centering
    \includegraphics[width=\columnwidth]{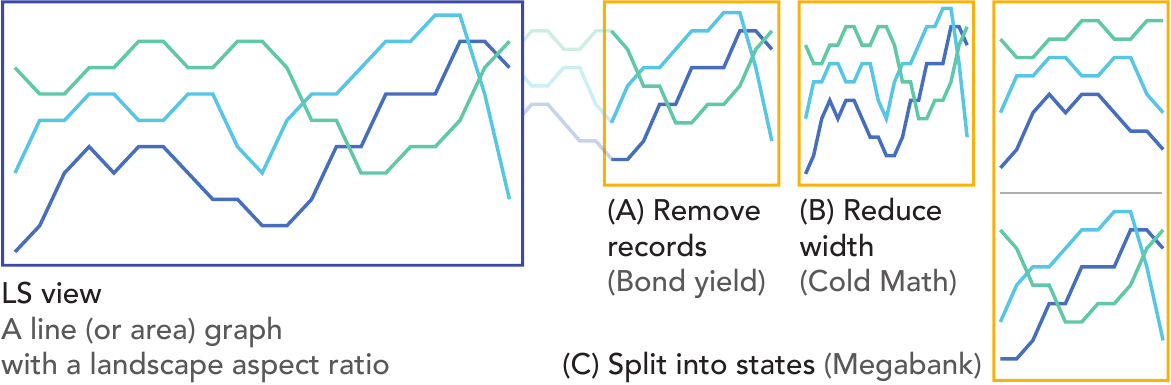}
    \caption{Motivating cases for trade-off analysis. From LS line or area chars with a wide landscape aspect ratio, (A) \textit{Bond Yield} removes records, (B) \textit{Cold Math} reduces width, and (C) \textit{Megabank} split states into panels for SS views.}
    \label{fig:tradeoff_example}
\end{figure}

\subsection{Method}\label{sec:study2method}
In analyzing strategies, we reflected on what problems seemed likely to have led to the use of the strategy. 
For example, in~\autoref{fig:tradeoff_example}, \textit{Bond Yield}, \textit{Cold Math} (E13), and \textit{Megabank} (E132) are line or area charts with a wide landscape aspect ratio. 
The author of \textit{Bond Yield} may have wanted to address graphical density by removing records, while the author of \textit{Cold Math} may have allowed changes to graphical perception to cope with the layout problem. The author of \textit{Megabank} may have compromised interactive sequencing to address interaction complexity.

We first identified and compared LS visualizations in our sample that shared similar design properties (\eg high cardinality, a wide aspect ratio, multiple views, interaction features) and noted differences in applying design strategies in these cases. Through discussion and iterative coding passes, we taxonomized problems the authors may have wanted to address. 
Similarly, we sought to code ways in which visualization messages are changed or lost in SS views by applying those design strategies. Where possible, we drew on existing literature on visualization perception and interaction to motivate losses and density problems to enhance our understanding. 
We also noted how authors may have attempted to compensate for such changes in messages. 
We captured our evolving understanding of trade-offs during this coding process in an affinity graph mapping design problems, strategies, changes in message, and compensation, and iterated on this graph several times, returning often to our sample. 

We summarize our results below but provide detailed characterizations of specific combinations of design choices manifesting trade-offs in our interactive gallery, each in terms of the underlying design problem prompting various design strategies and the downstream consequences of applying them. 
\subsection{Density Challenges}\label{sec:forms_density}

\bstartnc{Graphical density} poses challenges for authors
when maintaining a large number of objects (\eg marks, labels, annotations) in SS views
results in higher information density.
A high information density may make it difficult to identify or perceive differences between data points (\eg overplotting; c.f.,~\cite{heer2009,correll2019}).

\bstartnc{Layout} challenges occur
when it is difficult to maintain the arrangement of bigger objects, 
such as individual visualizations, legends, or interaction widgets on an SS display.
For example, fixed position elements, such as an overview and an interaction widget,
may consume a larger proportion of the screen space on SS.
Proportional rescaling of a visualization may overflow a single scroll height on SS, diminish the perceptibility of differences between values on a vertical scale,
or decrease the impact of the visualization on the viewer's impressions by reducing its relative size.

\bstartnc{Interaction complexity} challenges occur when an interaction feature is not feasible on SS because it requires immediate rendering of numerous graphical objects (computing power) and/or more precise manipulation than is possible on SS (\eg~due to the fat-finger problem~\cite{Lee2012}).

\subsection{Forms of Message Loss}\label{sec:forms_loss}
\bstartnc{Loss of information} stems from the fact that one of the easiest ways to reduce graphical density or interaction complexity is to omit some information (\eg \textit{remove fields}, \textit{remove annotations}). 
However, removing data, encodings, panels, or annotations
may reduce the viewer's ability to get certain intended takeaways of a visualization, and make certain comparisons. This problem can happen when we remove explicit interpretations of data provided via text, overview views, distributional information, or information in the form of records or fields. 
For example, when authors aggregate data to adjust graphical density for a smaller screen, the viewer can no longer make inferences about the distribution of aggregated variables unless the author takes specific steps to encode distribution through summary marks (\eg error bars), changes encodings, or adds details-on-demand. 
When a fixed overview visualization is removed on the SS view, viewers may take a longer time to explore the visualization~\cite{burigat2013,HORNBAEK2011509}.

\bstartnc{Loss of interaction} refers to loss of information that is available through interacting with a visualization (\eg~sorting data in a visualization by the viewer's criteria of interest).
For example, when it is difficult to render a feature immediately on SS browsers, authors may remove it, or split states that a viewer previously reached through interaction into static panels. 
However, such changes may result in loss of other states that users can find by interacting with the LS view.

\vspace{3mm}
\bstartnc{Reduced discoverability} refers to how using toggles and tabs to maintain information at a more appropriate density for a smaller screen can reduce viewers' abilities to find the information. 

\bstartnc{Reduced concurrency of elements} results from how the reduced screen space on SS devices often leads authors to choose to serialize elements, such that serialized elements are no longer visible within a single scroll height of SS display. 
This can hamper comparisons across visualizations. 
Transposing \textit{x}- and \textit{y}-axes in a two dimensional view to better fit the portrait layout of SS
can also lead to a visualization that is too long to fit within a single scroll height, hampering the viewer's ability to compare different values and assess high-level trends.

\begin{figure}[b]
    \centering
    \includegraphics[width=\columnwidth]{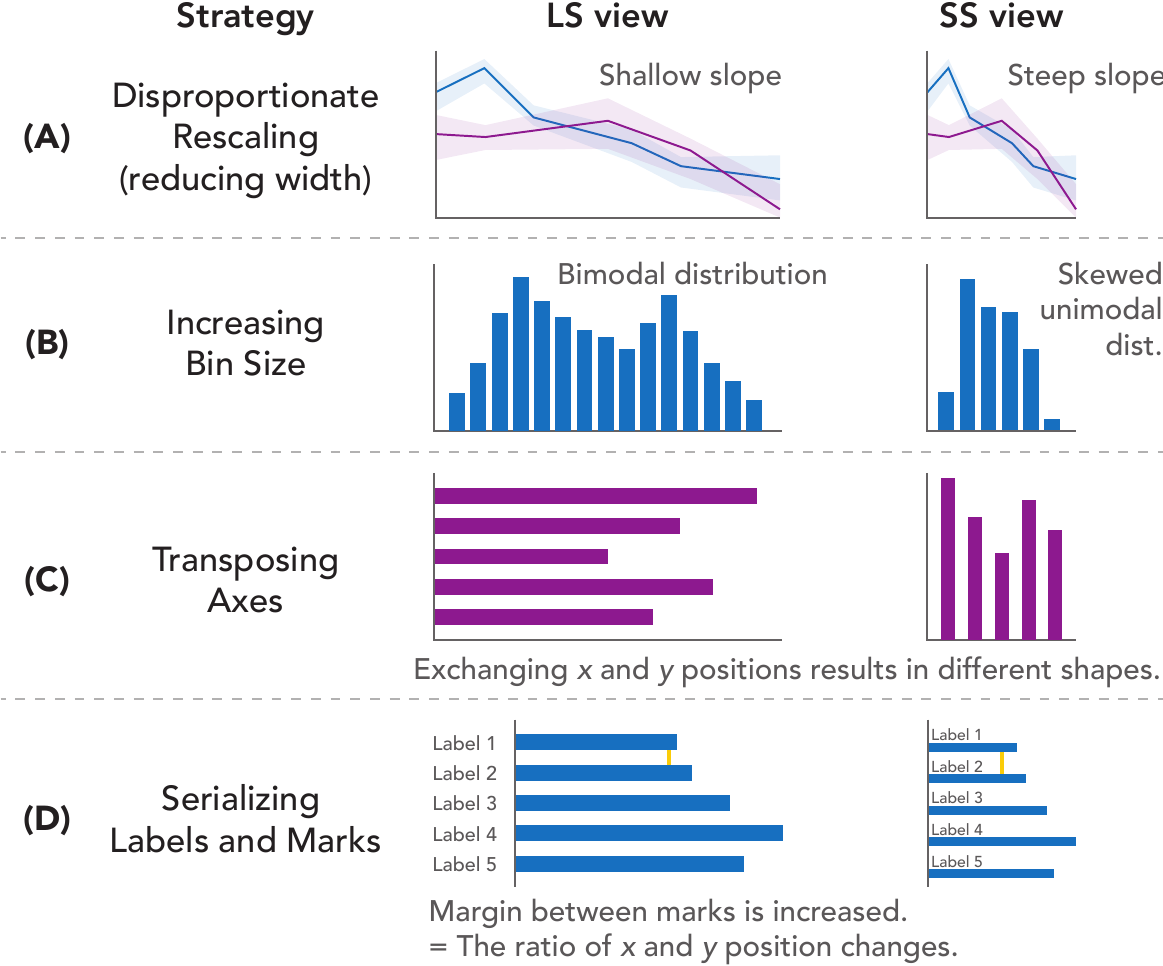}
    \caption{Examples of how transforming visualizations to fit narrower screen dimensions can change graphical perception.}
    \label{fig:aspectratio}
\end{figure}

\bstartnc{Changes in graphical perception} can result from responsive strategies like disproportionate rescaling, increasing bin size in aggregated views, transposing axes, or serializing labels and marks, as illustrated in \autoref{fig:aspectratio}.
Previous graphical perception studies \cite{Cleveland1985,talbot2012} have focused on slope perception; 
however, the graphical perception of other distributional information, such as dispersion or uncertainty may be affected by an aspect ratio change if they are encoded by position, length, or size channels.
More generally, changing an encoding channel (\eg~position to color value, E213), transposing axes (\eg~\textit{U.S. Cabinet}), or changing the range of a mapping (\eg~\textit{serialize label-marks}) for an SS view is likely to prompt different impressions among readers relative to graphical perception in the LS view.

\subsection{Complexity of the Density-Message Trade-off}\label{sec:tradeoff_complexity}
One reason navigating the trade-off between preserving information density and preserving message may be challenging is that what constitutes a message is often nebulous and rarely rigorously defined by an author. 
Instead, authors may experiment with different alternatives, relying on gisting and intuition to know when a change in design has hampered their goals.
To evaluate message preservation between responsive alternatives, authors need to be sensitive to changes in visual attributes.
For instance, a trend estimated on the same data (\eg~by regression) is invariant between responsive alternatives, while the visually implied trends may look different if those alternatives have changed aspect ratios, aggregation levels, or encodings.

The relationship between information and message preservation is also not always a direct mapping.
In some cases, removing information or interactivity may strengthen a message, if that information was not critical to it. 
For example, if being able to detect an anomaly is an intended goal for viewers, then removing records that are not anomalies and do not significantly change the distribution should not affect their ability. 
If, however, the author prioritizes trend recognition, they could aggregate data in a way that preserves the slope of a best fit line or other properties like clusters, while obtaining a more appropriate degree of graphical density.
To grapple with density-message trade-offs, authors must therefore think carefully about what they want to convey as takeaways and reason about the relative importance of different takeaways.
In doing so, authors need to compare their ranking of different takeaways to what they perceive as changes in emphasis on those takeaways under different responsive alternatives.
These complexities lead us to motivate formalizing a notion of messages in visualization to take steps toward providing automated support for exploring design alternatives.
\section{Discussion}
Our characterization of design strategies, patterns, and trade-offs in responsive transformation of communicative visualization informs visualization research and practice in several ways.
The design space implied by our results, which is captured by our design gallery, can help authors of responsive visualizations explore a larger space of design strategies. 
This more comprehensive coverage of the design space is useful to authors, who currently must rely on resources that describe strategies at a high level~\cite{hoffswell2020} or that provide technical documentation on a few, often more common strategies (\eg~responsive layout). 
Our results can also inform the design strategies that manual authoring tools (\eg~\cite{hoffswell2020}) or machine learning-based approaches (\eg~\cite{wu2020mobilevisfixer}) for responsive design support.
While this design space may not capture all responsive visualization strategies, we intentionally sought a relatively diverse sample spanning communication-oriented visualizations from the media, organizational reporting, and designerly interactive visualizations available online. 
Future work might extend the taxonomy we describe with more strategies as design innovations occur. 
Additionally, while we described our strategies from a default perspective of transitioning a LS view for smaller screens, the strategies we describe can be understood from the opposite, SS to LS direction. 

Many design problems are characterized by the negotiation of trade-offs. 
By outlining key trade-offs in responsive visualization in terms of what types of information are ``lost'' by certain design strategies concerning information density, our work aims to deepen understanding of the unique challenges. Our initial trade-off analysis provides only a first step to this larger goal. 
As people consume increasing amounts of information in multi-device environments, research effort around how to ensure that a set of visualizations capture the same ``takeaways'' despite design differences will become more important.

Moreover, the question of when a visualization preserves a message is integral to the process of designing visualizations for communication more broadly, as authors implicitly consider message preservation whenever they try out alternative designs. 
Currently, these judgments remain mostly subjective. 
A rich space for future research is to develop automated algorithms for predicting when two visualizations deliver the same ``message,'' operationalized as how well they support various tasks.  
Such attempts might use linear programming from human judgments as in GraphScape~\cite{kim2017graphscape}, or use deep learning models (\eg~\cite{haehn2018evaluating}) or graphical statistical inference models (\eg~\cite{hullman2019authors}) trained with human judgments. 
Formal approaches to capturing a visualization's message may be useful in visualization design applications beyond responsive visualization, like simplification of content for different audiences.

\subsection{Toward Responsive Visualization Recommendation}
Our analysis naturally motivates the development of recommender systems for responsive visualization that leverage the primary density-message trade-off we identified.
Given that strategies aimed at adjusting information density can lead to information loss in views for other screen sizes, it is important for authors to carefully consider which responsive view in the space of possible views achieves appropriate density while maintaining intended impressions or messages.
Authors may fixate on a design~\cite{duncker1945,bigelow2014reflections}, such as an LS view that they have already thoroughly considered. They may also stick with a design presumably to avoid time-consuming design iterations for responsive alternatives~\cite{hoffswell2020}.
That the same design specification does not lead to the same responsive transformation implies responsive visualization is a wicked problem where no single optimal solution may always exist~\cite{buchanan1992,rittel1973}.
A lack of guidelines, combined with the reasons above, motivate tools that can help authors explore, perceive, and reason about the space of possible responsive alternatives given an original view.

A recommender approach requires formulating responsive visualization design as a search problem from a source view to target views (\eg~from LS to SS views).
We assume a scenario in which given a source view, a recommendation system populates a set of possible target designs ranked by how well they address density-message trade-offs.

An automated recommender approach entails several requirements that future work might consider.
The first step to such an automated approach is to \textbf{encode responsive design strategies} to allow for a wider search space for alternative responsive views. 
Declarative grammars for visualization recommendation would be useful to generate a search space by formalizing design knowledge regarding responsive visualization as well as conventional guidelines 
(\eg~well-formedness~\cite{moritz2018formalizing}, effectiveness~\cite{mackinlay1986automating}).
In particular, a constraint-based approach that formalizes design knowledge as constraints, such as Draco~\cite{moritz2018formalizing}, can encode our design patterns as constraints (\eg~aggregating data with high cardinality, fixing tooltip position for smaller screens).
Our representation of Actions as invertible functions with input and output states would be a useful schema in decomposing and formalizing design patterns (\eg~\textit{increase bin size} as \texttt{from: bin(size=15, field=A, ...)} $\rightarrow$ \texttt{to: bin(size=25, ...)}).
Machine learning based approaches could also encode design strategies as multiple classification problems (\eg~a model that predicts whether each design strategy is applicable to a given source visualization).

Considering the complexity of density-message trade-offs (\autoref{sec:tradeoff_complexity}), future work should pursue ways to \textbf{operationalize message preservation} so that authors can better compare the relative importance of different messages under different alternatives.
While it is not always easy to define visualization messages because of their implicitness, subjectivity, and domain specificity~\cite{North2006}, prior approaches to insight-based visualization recommenders~\cite{tang2017topk,burns2013,cui2019,demiralp2017,Srinivasan2019} have estimated visualization messages or insights based on analytic tasks (\eg~\cite{amar2005,Brehmer2013}) such as estimating correlation and mean, finding data ranges, strength of trend (regression coefficients) from data. 
Research on how visualizations communicate messages or narratives may also be informative about the kinds of goals authors often have when designing charts~\cite{Segel2010,Hullman2011}.

\section{Conclusion}
More and more users are viewing and interacting with visualizations on small screen devices. 
We contribute a detailed characterization of 76 responsive visualization design strategies organized by their Targets and the Actions applied. 
We find that while simple transformations like rescaling a visualization are most frequently used in a sample of 378 pairs of LS and SS communicative visualizations, many authors are also exploring alternative strategies that appear aimed at adjusting graphical density to be appropriate for a smaller screen while also trying to maintain the message of a visualization. 
We articulate how design challenges stemming from density-message trade-offs arise in responsive visualization design, threatening the user's ability to do various visual inference tasks. 
We discuss the implications of our findings for existing approaches around responsive visualization and outline potential requirements for an automated recommender.
Our work contributes guidance for practitioners seeking to develop responsive visualizations and researchers interested in better understanding, and designing systems for, responsive visualization authoring.

\bibliographystyle{eg-alpha-doi}
\bibliography{egbibsample}


\newpage

\end{document}